\UseRawInputEncoding
\documentclass[11pt]{article}
\usepackage[symbol]{footmisc}
\def\correspondingauthor{\footnote{Corresponding author.  }}
\usepackage[left=2cm,top=2.5cm,right=2cm,bottom=2.5cm]{geometry}
\usepackage{amsmath, amssymb}

\usepackage{graphicx}
\usepackage{graphics}
\usepackage{epstopdf}
\usepackage{subfigure}
\usepackage{amsfonts}
\usepackage{sectsty}
\usepackage{sectsty}
\usepackage{hyperref}
\usepackage{cite, multirow,hhline}

\begin{document}
	\begin{center}
	\large{\bf{Cosmological dynamics of $f(R)$ models in dynamical system analysis}} \\
	\vspace{5mm}
	\normalsize{Parth Shah$^1$ and Gauranga C. Samanta$^{2,}{}$\correspondingauthor{}, }\\
	\normalsize{$^{1}$ Department of Mathematics, BITS Pilani K K Birla Goa Campus, Goa, India\\
	$^{2}$ P. G. Department of Mathematics, Fakir Mohan University, Odisha, India}\\
	\normalsize {parthshah2908@gmail.com \\gauranga81@gmail.com }
\end{center}
	
\begin{abstract}
\noindent
In this work we try to understand the late time acceleration of the universe by assuming some modification in 
the geometry of the space and using dynamical system analysis. This technique allows to understand the behavior of 
the universe without analytically solving the field equations. We study the acceleration phase of the universe and 
stability properties of the critical points which could be compared with observational results. We consider an 
asymptotic behavior of two particular models  $f(R) = R - \mu R_{c} \frac{(R/R_c)^{2n}}{(R/R_c)^{2n} + 1}$  
and $f(R) = R - \mu R_{c} \left[ 1 - (1 + R^2/R_{c}^2)^{-n} \right]$ with $n,\mu, R_c >0$ for the study. 
As a first case we fix the value of $\mu$ and analyzed for all $n$. Later as second case, we fix the value of 
$n$ and calculation are done for all $\mu$. At the end all the calculations for the generalized case have been shown and 
results have been discussed in detail.
\end{abstract}

%\textbf{AMS Classification number}: 83F05\\

\textbf{Keywords}: Dark energy, Modified gravity theory, Dynamical system analysis \\
%\textbf{Mathematics Subject Classification Codes:} 83C05; 83C15; 83F05.

\section{Introduction}
More than a century ago, Einstein proposed his theory of general relativity (GR) which revolutionized the idea of gravity. 
It is a theory that acted as a powerful tool in our pursuit to understand the universe. With the passage of time and with 
extensive research, snags in the theory started showing up. The major blow came towards the end of the last century when 
the accelerated expansion of the universe \cite{perlmutter, riess} was discovered which left GR inconsistent at the 
cosmological distances. Since then researchers have started looking for the alternative techniques such as modified 
gravity and dark energy to incorporate this accelerated expansion in the theory of gravity. While the former deals with 
the geometry of space-time, the latter is concerned with the matter content of the universe. These are two different
approaches to understand the late time acceleration of the universe. Extensive reviews in modified gravity can be
found in the Refs. \cite{nojiri, nojiri1, capozziello}. Many of such theories aim at modifying the linear function of
scalar curvature $R$ from its special form in GR to a more generic form. $f(R)$ gravity is one such attempt where the
gravitational lagrangian of GR, $\mathcal{L}_{GR} = R$ is replaced by an analytic function of $R$ i.e. 
$\mathcal{L}_{f(R)} = f(R)$. Choosing a suitable function for $f(R)$, one can explore the non-linear effects of 
the scalar curvature on the evolution of the universe. Extensive reviews in $f(R)$ gravity can be found in the refs
\cite{thomas, felice}. Viability of $f(R)$ dark energy models have been studied in ref. \cite{amendola}, 
where the $f(R)$ models with a power law of $R$ has been ruled out. Author of ref. \cite{thomas1} studied 
the interplay between $f(R)$ theories and scalar-tensor theories via the Palatini formalism. Formation of 
large scale structure in $f(R)$ gravity was studied in ref. \cite{song}. A reconstruction scheme for $f(R)$ 
theories was explored in ref.\cite{nojiri2}. Various other studies related to $f(R)$ gravity can be found in 
\cite{shah,odintsov,bamba,capozziello1}. There are many ways other ways to modify the general theory of relativity. 
Some of the other alternatives are Scalar Tensor Theory\cite{bergmann, faraoni, nordtvedt, wagoner, starobinsky4, 
starobinsky5, starobinsky6, starobinsky7}, Brans Dicke theory \cite{brans,dicke}, Gauss Bonnet theory \cite{nojiri3},
$f(T)$ gravity \cite{bamba1, bamba3, bamba4, bamba5, bamba6, bamba7} $f(R,T)$ gravity \cite{harko,bamba9}, 
$f(R,G)$ gravity \cite{bamba8}, Lovelock gravity \cite{lovelock, padmanabhan}. $f(R)$ gravity has been described in 
detail in \cite{amendola1, Romero, Hu, Starobinsky2007, salvatore1, Nojiri2011, bamba2, motohashi, motohashi1, starobinsky, starobinsky1, 
starobinsky2, starobinsky3}. Subsequently many  authors \cite{Bamba2014, Bamba2015, Nojiri2017, Yousaf2017, Godani, 
Odintsov2018, Capozziello2018, Odintsovsd} studied cosmological models from various aspects in modified gravity.

\noindent
It is very difficult to find out the analytic as well as numerical solution in general as well as modified theory of 
gravity. This is due to the fact that field equations contains nonlinear terms and that leads to difficulty in
comparison with observations. One such method to avoid this problem and to study the dynamical behavior of these 
equations is the dynamical system analysis. Many authors have used this approach in cosmology \cite{Roy, Odintsov, 
Odintsov1, Hohmann, Carneiro, Bhatia, Santos, Bamba, parth}. This methods aims at finding the numerical solution of 
the system which helps to understand the qualitative behavior of the system. First we obtain the critical points of 
the given set of first order differential equations. Then we apply the linear stability theory which is to linearise 
the system near the critical point in order to understand the dynamics of the whole system. In this theory only the 
first partial derivates is considered which corresponds to the Jacobian matrix in vector calculus and is also 
referred as stability matrix. The eigenvalues of the Jacobian matrix contain the information about the stability 
of particular critical point. Although this theory fails for non-hyperbolic points for which Lyapunov's method and 
central manifold theory are used often.   Application of dynamical systems analysis to cosmology has been discussed
in these books \cite{Ellis} and \cite{Coley}.

\noindent
In the present work we analyze the stability and acceleration phase of the cosmological model of the universe 
which is assumed to have modification in its geometry part, i.e. $f(R)$ gravity. We begin with understanding the
Metric formalism and Palatini formalism of $f(R)$ gravity. This work has been carried out in Metric formalism. 
In this work, we consider the universe to be composed of matter and radiation with no interaction between them. 
We begin by considering two different $f(R)$ models and show that their asymptotic behavior is same. 
We study the existence of stability phase and acceleration era for various cases by calculations  and plots. 
We conducted this study due to the unique property of two different $f(R)$ models to be an asymptote to a same model. 
Also we have studied the stability and acceleration based on the all possible values of parameters $\mu$ and $n$ for 
these models. In section 2 brief review of metric formalism of $f(R)$ theory is discussed, stability analysis and 
acceleration phase analysis of the model is done is section 3. Section 4 contains results, conclusions and future 
possible work.

%We start our analysis by considering our universe to be filled with dust (ordinary baryonic matter alongside dark matter) and radiation which are assumed to not interact with each other. Then we bring the simplest form of dark energy as cosmological constant ($p = -\rho$) into this model. In which we investigate the impact of  both linear and nonlinear interaction. By studying the behaviour of all these interactions, we show the existence of unstable and stable phases of the universe. This

\section{Metric $f(R)$ gravity}

As it is known that this theory comes as a straightforward generalization of the Lagrangian with matter part in 
the Einstein-Hilbert action,
\begin{equation}
S_{EH} = \frac{1}{2 \kappa} \int d^4x \sqrt{-g} R + S^{(m)}
\end{equation}
where $\kappa = 8\pi G$, $R$ is Ricci Scalar, $g$ is determinant of metric, $g_{\mu\nu}$ = diag$(-1,a^2(t),a^2(t),a^2(t))$
and $a(t)$ is scale factor to become a general function of $R$, i.e.
\begin{equation}{\label{2}}
S = \frac{1}{2 \kappa} \int d^4x \sqrt{-g} f(R) + S^{(m)}
\end{equation}
where
$f(R)$ is a non-linear function of its argument and $S^{(m)}$ is the matter part of the action.
%There are two variational principles that one can apply to the Einstein-Hilbert action in order to derive Einstein's equations: the standard metric variation and Palatini variation. In the Palatini variation the metric and the connection are assumed to be independent variables and one varies the action with respect to both of them, under the important assumption that the matter action does not depend on the connection. The choice of the variational principle is usually referred to as a formalism, so one can use the terms metric which is a second order formalism and Palatini which is a first order formalism. We here note that both the variational principles lead to the same field equation for an action whose Lagrangian is linear in $R$ but this is no longer true for a more general action. Therefore, it is clear that there will be two version of $f(R)$-gravity, according to which variational principle or formalism is used. Indeed this is the case: $f(R)$-gravity in the metric formalism is called metric $f(R)$-gravity and $f(R)$-gravity in the Palatini formalism is called Palatini f(R)-gravity. This work uses the standard metric or the second order  variation approach for all discussions.
Variation of this action in standard metric formalism with respect to the metric $g^{\mu \nu}$ gives
\begin{equation}{\label{3}}
F(R) R_{\mu \nu} - \frac{1}{2} f(R) g_{\mu \nu} + [ \square g_{\mu \nu} - \nabla_{\mu} \nabla_{\nu}] F(R) = \kappa T_{\mu \nu}
\end{equation}
where, $F(R)$ (also denoted $f_{,R}$) is $\frac{\partial f}{\partial R}$ and as usual,
\begin{equation}{\label{4}}
T_{\mu \nu} = \frac{-2}{\sqrt{-g}}\frac{\delta S_m}{\delta g^{\mu \nu}}
\end{equation}
where, $\nabla_{\mu}$ is the covariant derivative associated with the metric and $\square \equiv \nabla^{\mu}\nabla_{\mu} $.\\
It can be noted from \eqref{3} that these are fourth order differential equations in the metric, since $R$ contains 
second order partial derivatives. Theory is reduced to GR when the action contains only $R$ since the last two terms of 
the left hand side vanishes. The trace of equation \eqref{3} is given by
\begin{equation}
3\square F(R) + F(R)R - 2f(R) = \kappa^2 T
\end{equation}
We also note that field equations could be written in the form of Einstein equations by moving the effective 
stress-energy tensor to the right hand side. Specifically, \eqref{3} can be re-written as
\begin{equation}
\begin{aligned}
G_{\mu \nu} &\equiv R_{\mu \nu} - \frac{1}{2}g_{\mu\nu}R \\
&= \frac{\kappa^2 T_{\mu\nu}}{F(R)} + g_{\mu \nu} \left[ \frac{f(R) - RF(R)}{2F(R)} \right]  + \left[ \frac{\nabla_{\mu}
\nabla_{\nu} - g_{\mu \nu} \square F(R)}{F(R)} \right]
\end{aligned}
\end{equation}
or
\begin{equation}
G_{\mu \nu} = \frac{k}{F(R)} \left( T_{\mu\nu} + T_{\mu\nu}^{(eff)} \right)
\end{equation}
This gives an effective stress-energy tensor which does not have the canonical form quadratic in the first 
derivatives of the field $f(R)$, but contains terms linear in the second derivatives. Also it was very essential 
to form set of conditions which are viable for $f(R)$ models in metric formalism. These conditions have been stated and 
discussed in \cite{amendola1,Romero,shah}.
%\begin{itemize}
%\item $f_{,R} > 0$ for $R \geq R_{0}$ (where $R_{0}$ is the Ricci scalar at present epoch and is positive). If the final attractor is a  This condition is required to avoid anti-gravity.
%
%\item $f_{,RR} > 0$ for $R \geq R_{0}$. This is required for local gravity tests \cite{dolgov, olmo, faraoni1, navarro} for existence of matter domination era \cite{amendola, amendola2} and for stability of cosmological perturbation \cite{carroll3, song, bean, faulkner}.
%
%\item $f(R) \to R - 2\Lambda$ for $R \gg R_{0}$. This is required for consistency with local gravity tests \cite{amendola3, hu, starobinsky2, appleby, tsujikawa2} and for presence of matter dominated era \cite{amendola}.
%
%\item $0 < \frac{R f_{,RR}}{f_{,R}} < 1$ at $- \frac{R f_{R}}{f} = -2$. This is required for the stability of the late-time de Sitter point \cite{amendola,muller,faraoni2}.
%\end{itemize}

\section{Stability Analysis}
In this paper, the following viable $f(R)$ models have been considered to explain cosmic acceleration using dynamical 
system techniques.
%In our current work, we analyze following models

\begin{equation}
\begin{aligned}
f(R) &= R - \mu R_{c} \frac{(R/R_c)^{2n}}{(R/R_c)^{2n} + 1}  \\
f(R) &= R - \mu R_{c} \left[ 1 - (1 + R^2/R_{c}^2)^{-n} \right]
\end{aligned}
\end{equation}
with $ n, \mu, R_{c} > 0$  \cite{amendola1, Hu, Starobinsky2007}. This model satisfies all the local gravity 
conditions and are considered to be a viable model to study the stability analysis of the universe. It is noted here 
in both these models the function $f(R)$ asymptotically behaves as $f(R) \rightarrow R - \mu R_c   
\left[ 1 - (R^2/R_{c}^2)^{-n} \right]$ for $R \gg R_c$. These models also satisfy $f(R=0)=0$, so the cosmological 
constant vanishes in the flat space time. These models are chosen as the property that two $f(R)$ models are asymptote 
to a single $f(R)$ model is very difficult to find.
For the flat FLRW space-time the Ricci scalar is given by:
\begin{equation}
R = 6(2H^2 + \dot{H})
\end{equation}
where $H$ is the Hubble parameter. We now construct a model of the universe filled with only matter and radiation 
and we assume no interaction between them i. e. the usual conservation equations $\dot{\rho_m} + 3H \rho_m =0$ 
and $\dot{\rho_r} + 4H \rho_r =0$. We also assume that matter is pressure-less i.e. $p_m = 0$. 
For this, the explicit form of field equations from equation \eqref{4} are
\begin{equation}\label{8}
\begin{aligned}
3FH^2 = \kappa^2(\rho_m + \rho _r) + \frac{FR - f}{2} - 3H\dot{F} \\
-2F \dot{H} = \kappa^2 (\rho_m + \frac{4}{3} \rho_r) + \ddot{F} - H\dot{F}
\end{aligned}
\end{equation}
Now, we would like to convert the above non-autonomous field equations to an autonomous systems by introducing 
the following dimensionless variables,

%Now introducing dimensionless variables,

\begin{equation}
x = - \frac{\dot{F}}{HF}, y = - \frac{f}{6FH^2}, z = \frac{R}{6H^2}, w = \frac{\kappa^2 \rho_r}{3FH^2}
\end{equation}
Without loss of generality, let $\kappa^2 = \frac{8 \pi G}{c^4} = 1$.
Then, the various density parameters would be,
\begin{equation}
\Omega_r = \frac{\rho_r}{3FH^2} = w, \Omega_m = \frac{\rho_m}{3FH^2} = 1 - x - y - z - w, \Omega_{GC} = x + y + z
\end{equation}
where, $\Omega_{GC}$ represents density parameter due to geometric curvature. From equation \eqref{8}, 
it is straightforward to derive following set of autonomous differential equations

\begin{equation}
\begin{aligned}
x' &= -1 - z - 3y + x^2 - xz + w \\
y' &= \frac{xz}{m} - y(2z - 4 - x) \\
z' &= - \frac{xz}{m}- 2z(z-2) \\
w' &= -2zw + xw
\end{aligned}
\end{equation}

\noindent
where, prime denotes derivative with respect to $\eta = ln a$ and
\begin{equation}
\begin{aligned}
m &\equiv \frac{dlnF}{dlnR} = \frac{R f_{,RR}}{f_{,R}} \\
r &\equiv - \frac{dlnf}{dlnR} = - \frac{Rf_{,R}}{f} = \frac{z}{y}
\end{aligned}
\end{equation}
From this, $R$ could be written as a function of $\frac{z}{y}$. We here note that $m$ is a function of $R$, 
so it follows that $m$ is a function of $r$, i.e. $m = m(r)$. From the calculations for the model, $f(R) = R - 
\mu R_c   \left[ 1 - (R^2/R_{c}^2)^{-n} \right]$ we deduce
\begin{equation}\label{15}
m = \frac{2n(2n+1)}{\mu^{2n}}(-r-1)^{2n+1}
\end{equation}
By linear approximation and substituting value of $r$, $m$ could be rewritten as:
\begin{equation}\label{16}
m = -\frac{2n(2n+1)}{\mu^{2n}}\left( 1 + (2n+1)\frac{z}{y} \right)
\end{equation}
We analyze these equations in two different cases.\\

\noindent
\textbf{Case A}: $n$ is unknown and $\mu = 1$.\\
By substituting \eqref{16} in autonomous differential equations, and assuming $\mu = 1 $ we get

\begin{equation}
\begin{aligned}
x' &= -1 - z - 3y + x^2 - xz + w \\
y' &= \frac{xyz}{2n(2n+1)[y+(2n+1)z]} - y(2z - 4 - x) \\
z' &= - \frac{xyz}{2n(2n+1)[y+(2n+1)z]} - 2z(z-2) \\
w' &= -2zw + xw
\end{aligned}
\end{equation}
also,
\begin{equation}
\omega_{eff} = -1 - \frac{2 \dot{H}}{3H^2} = - \frac{1}{3} (2z-1)
\end{equation}
There are 8 real critical points of this system. We will now do the detailed stability and acceleration 
analysis for all the points. Plot legends 1, 2, 3, etc have been used to denote different eigenvalue of 
particular critical point.\\

\noindent
$\mathbf{P_1}: (-4,5,0,0)$. $\Omega_m = 0, \Omega_r = 0, \omega_{eff} = \frac{1}{3}$.
Eigenvalues of this critical point are $-3, -4, -5$ and $ \frac{2(4n^2 + 2n -1)}{n(2n+1)}$.  
The eigenvalues of this critical point is converging to 4. This point is stable for only the small range of 
$n$ $(n < \frac{\sqrt{5} - 1}{4})$. Since the value of $\omega_{eff}$ is positive, acceleration for this model is 
could not possible.

\begin{center}
\includegraphics[scale=0.8]{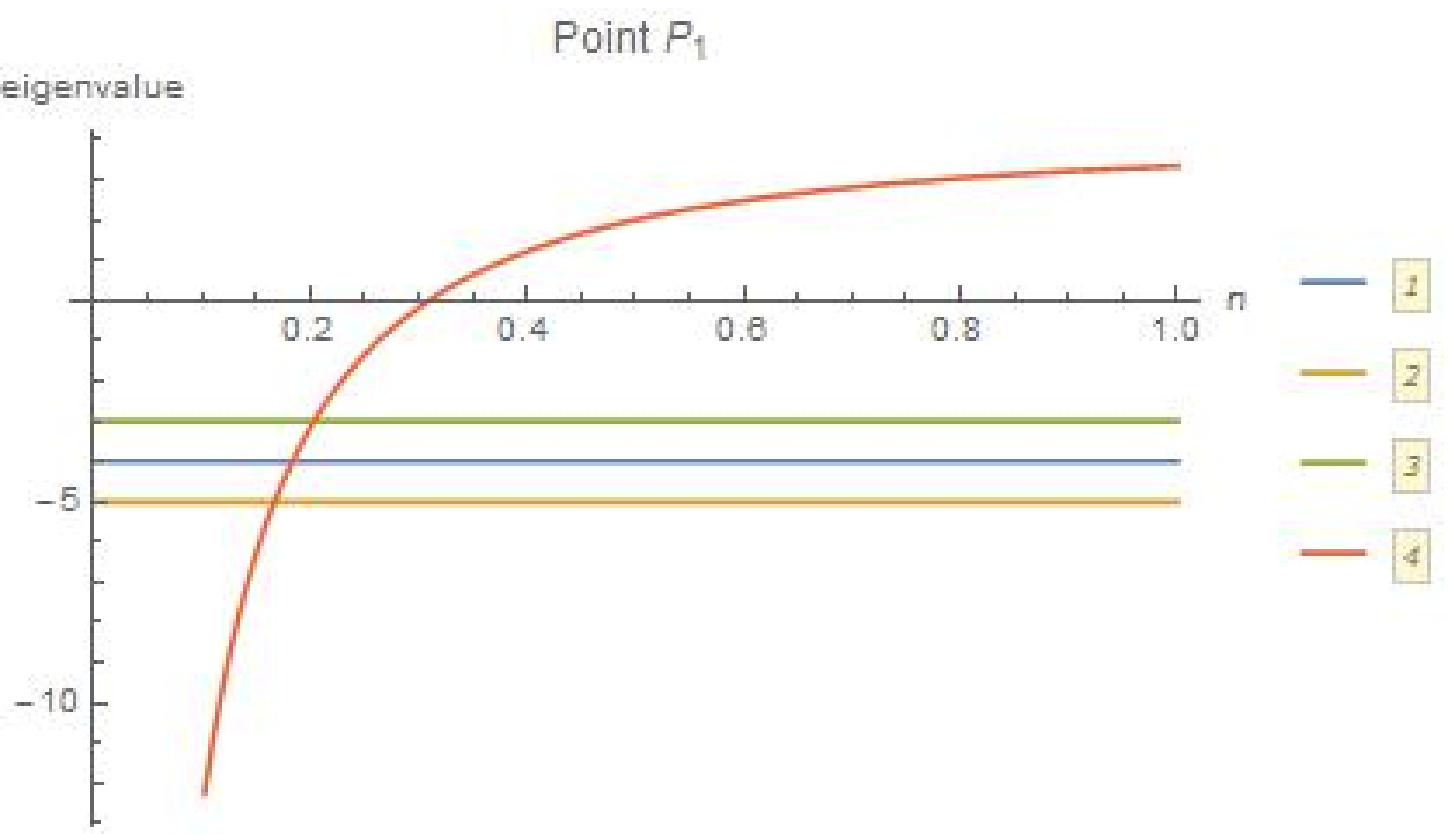}
\end{center}

\noindent
$\mathbf{P_2}: (0,-1,2,0)$. $\Omega_m = 0, \Omega_r = 0, \omega_{eff} = -1 $.
We here observe that real part of all the eigenvalues are negative for $0.152379 \leq  n < 0.19908 $. 
Hence this point is spiral stable for a small range. Apart from that $\omega_{eff}$ is negative, hence this 
point gives acceleration. This point is completely dominated by geometric curvature as $\Omega_{GC} = 1$. Since there
is no matter or radiation component in this point, this point could be considered to be responsible for late time 
acceleration of the universe.

\begin{center}
\includegraphics[scale=0.8]{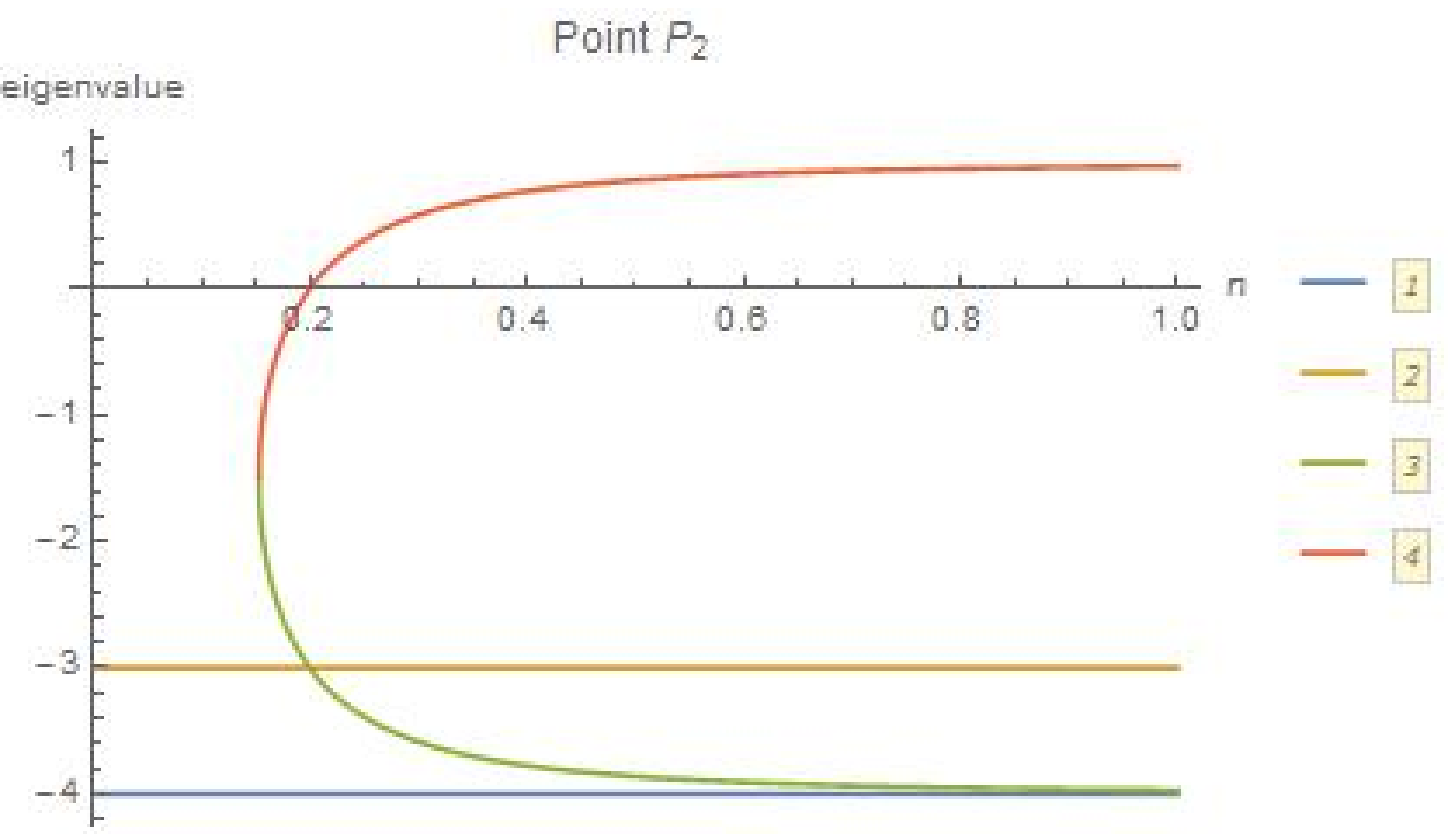}
\end{center}

\noindent
$\mathbf{P_3}: (-1,0,2,0)$. $\Omega_m = 0, \Omega_r = 0 ,\omega_{eff} = -1$.
Eigenvalues of this critical point are:$-5, -4, -4,$\\ $\frac{-1-2n-8n^2-8n^3}{2n(1+2n)^2}$. 
This model is stable for $n > \frac{1}{24} \left( -8 + \sqrt[3]{(928 - 96 \sqrt{93}} + 2 * 2^{2/3} 
\sqrt[3]{29 + 3\sqrt{93}} \right)$. This model could provides acceleration since $\omega_{eff}$ is negative. 
Also we note here that this critical point is a `stable proper node' as two distinct eigenvectors arises for 
repeated eigenvalues. Like $P_2$, this point is also completely dominated by geometric curvature but for a large 
range of $n$ value which is evident from the plot below.

\begin{center}
\includegraphics[scale=0.8]{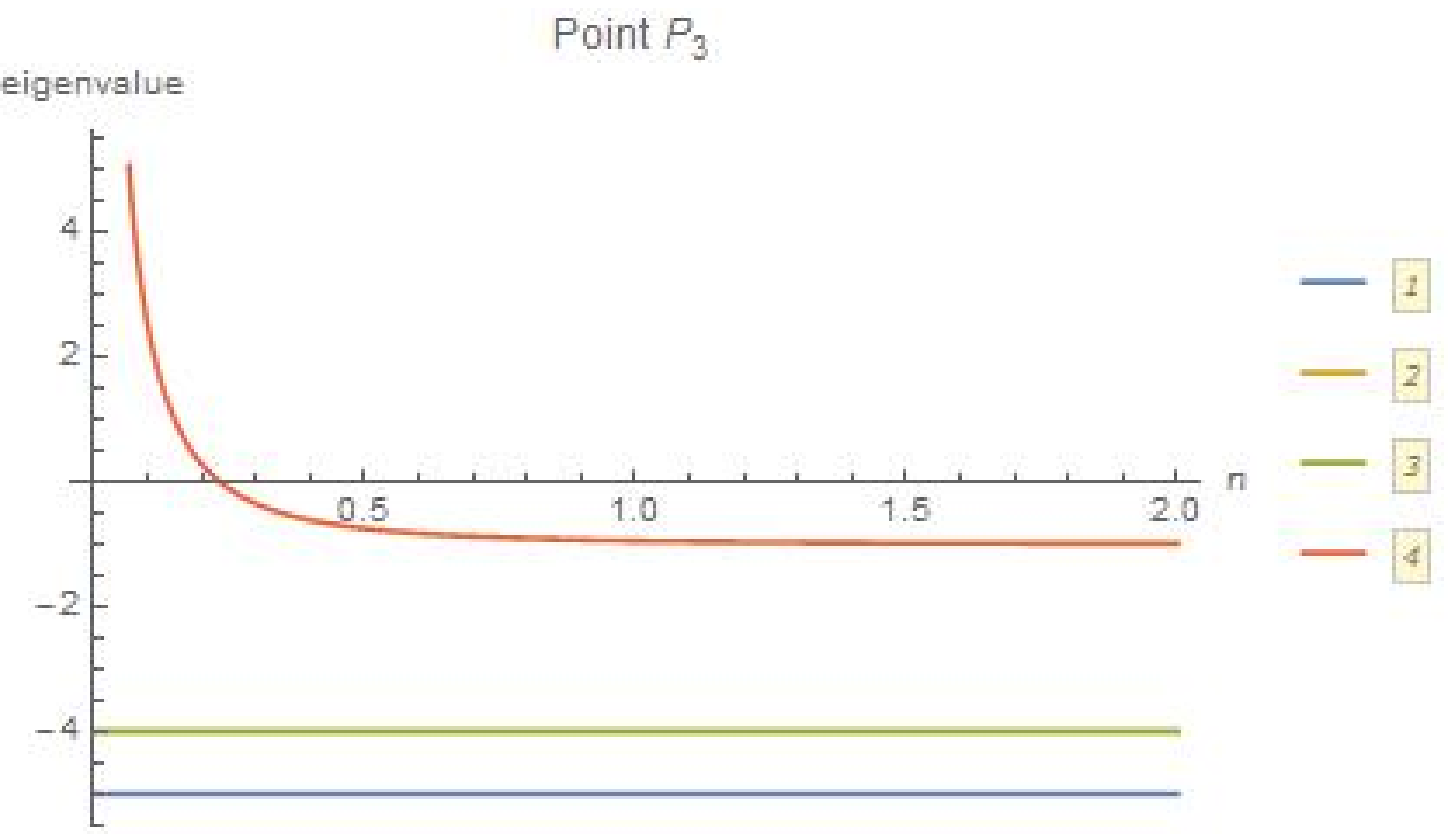}
\end{center}

\noindent
$\mathbf{P_4}: (3,0,2,0)$. $\Omega_m = -4, \Omega_{r} = 0 ,\omega_{eff} = -1$. Eigenvalues of this critical 
point are: $-4, -1, 4,$ $\frac{3(1+2n+8n^2+8n^3)}{2n(1+2n)^2}$. It is clear that since one eigenvalue is positive, 
this point is not stable and acceleration occurs for this point.\\

\noindent
$\mathbf{P_5}: (4,0,2,-5)$. $\Omega_m = 0, \Omega_{r} = -5 ,\omega_{eff} = -1$. Eigenvalues of this critical point are: 
$-4, -1, 5, \\ \frac{2(1+2n+8n^2+8n^3)}{2n(1+2n)^2}$. Again here one eigenvalue is positive hence this point is not 
stable and acceleration occurs for this point as $\omega_{eff} = -1$.\\

\noindent
$\mathbf{P_6}: \left( - \frac{12(n^2+2n^3)}{-1+2n+4n^2}, \frac{(-1+2n+8n^2+8n^3)(1-2n+8n^2+24n^3)}{2(-1+2n+4n^2)^2},
\frac{-1+2n-8n^2-24n^3}{2(-1+2n+4n^2)},0 \right)$.\\ $\Omega_m = \frac{1-4n-18n^2+16n^3+104n^4+16n^5-96n^6}{(-1+2n+4n^2)^2},
\Omega_r = 0, \omega_{eff} = \frac{4n^2 + 8n^3}{-1+2n+4n^2}$.
From the value of $\omega_{eff}$ we see that the acceleration for this model occurs from $ \frac{1}{72} 
\left( -16 + \sqrt[3]{7136 - 288 \sqrt{597}} + 2^{\frac{5}{3}}\sqrt[3]{223 + 9 \sqrt{597}} \right)  
< n < \frac{1}{4}(\sqrt{5} - 1)$. Also the real part of all the eigenvalues is negative in few regions for $0 < n < 1.5$ 
as shown in the figure. So for this model, the stability and acceleration can occur simultaneously but it is in very small
region. Since the value of $\Omega_m$ for this point is non zero, this point can be used for the matter-dominated epoch. 
This point does not govern acceleration and stability for a long duration which makes it a better candidate for matter 
dominated point.
\begin{center}
\includegraphics[scale=0.8]{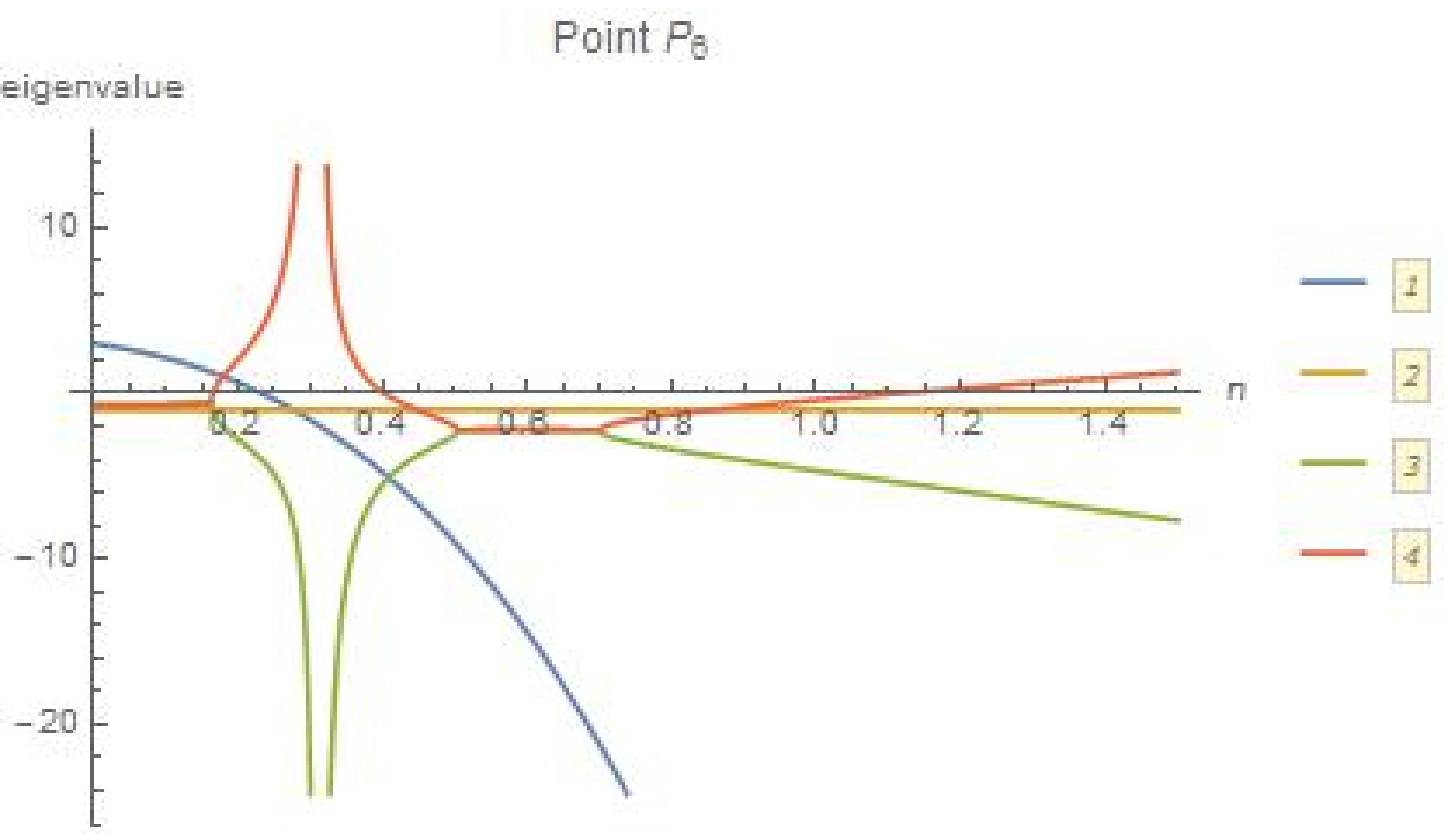}
\end{center}

\noindent
$\mathbf{P_7}: \left( -\frac{16 \left(2 n^3+n^2\right)}{4 n^2+2 n-1}, \frac{8 \left(16 n^6+24 n^5+12 n^4-n^2\right)}
{\left(4 n^2+2 n-1\right)^2}, -\frac{8 \left(2 n^3+n^2\right)}{4 n^2+2 n-1}, \frac{-128 n^6+112 n^4+16 n^3-20 n^2-4 n+1}
{\left(4 n^2+2 n-1\right)^2}\right)$. Two of four eigenvalues of this point are 1 and $-4(-1 + 2n + 8n^2 + 8n^3)$. 
The second eigenvalue would be negative for $n > \frac{1}{24} \left( -8 + \sqrt[3]{(928 - 96 \sqrt{93}} + 2 * 2^{2/3} 
\sqrt[3]{29 + 3\sqrt{93}} \right)$ i.e. $n > 0.2328$. Also the real part of third and fourth eigenvalue are negative for
small n. Hence we do not get any region were all the other three eigenvalues are positive. Hence this would be a saddle 
point due to opposite signs of eigenvalues. Moreover, this point could provide acceleration for,
$\frac{1}{48}\left( -12 + \sqrt[3]{3456 - 192 \sqrt{321}}  + 4 \sqrt[3]{3 (18 + \sqrt{321}} \right) 
< n < \frac{1}{4} (\sqrt{5} - 1)$ which is again very short lived.\\

\noindent
$\mathbf{P_8}: \left( - \frac{2(-1+2n+12n^2+16n^3)}{1-2n+8n^3},  \frac{(-1+2n+8n^2+8n^3)(-1+2n+24n^2+40n^3)}{4n^2
(1-4n^2+8n^3+16n^4)}, \frac{-1+4n+24n^2-16n^3-176n^4-160n^5}{4n^2(1-4n^2+8n^3+16n^4)}, 0 \right)$.
The $\omega_{eff}$ here suggests that the model does not give any acceleration for $n > 0$. 
But this model becomes stable in $n > \sqrt{96 n^6 - 16 n^5 - 104 n^4 - 16 n^3 + 18 n^2 + 4 n - 1}$ i.e near
$n = 1.11569 $, which is evident from the plot.

\begin{center}
\includegraphics[scale=0.8]{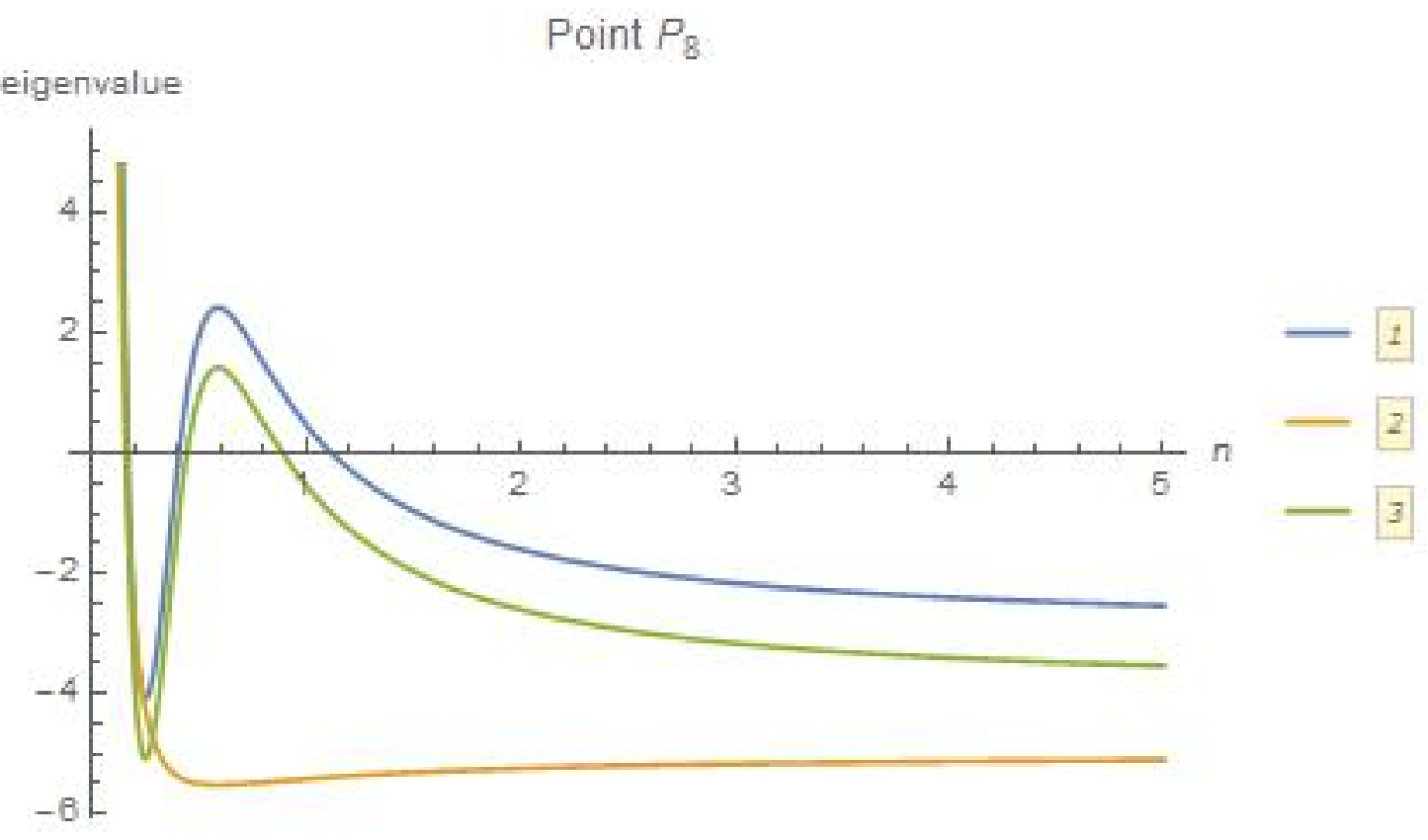}
\end{center}

\noindent
\textbf{Case B}: $\mu$ is unknown and $n = 1$.\\
As a second case, we follow a different approach, i.e. by evaluating these equations for different values of 
$n$ ($n > 0$) and $\mu$ being an unknown variable and study the dynamics of the model. The set of autonomous differential
equations for $n=1$ would be as follows:

\begin{equation}
\begin{aligned}
x' &= -1 - z - 3y + x^2 - xz + w \\
y' &= - \frac{xy^3z\mu^2}{6(y+z)^3} - y(2z - 4 - x) \\
z' &= \frac{xy^3z\mu^2}{6(y+z)^3} - 2z(z-2) \\
w' &= -2zw + xw
\end{aligned}
\end{equation}
There are 5 real critical points of this system. Their detailed stability and acceleration analysis would be done.

\noindent
$\mathbf{P_9}: (-4,5,0,0)$. $\Omega_m = 0, \Omega_r = 0, \omega_{eff} = \frac{1}{3} $. Eigenvalues 
are -5, -4, -3 and 4 - $\frac{2\mu^2}{3}$. This point does not show any acceleration but stability occurs 
for $0 < \mu < \sqrt{6}$. This point is similar to $P_1$.\\

\noindent
$\mathbf{P_{10}}: (0,-1,2,0)$. $\Omega_m = 0, \Omega_r = 0, \omega_{eff} = -1 $. Eigenvalues are -4, -3, 
$\frac{1}{6} \left(-9 - \sqrt{3} \sqrt{75 - 8\mu^2} \right) $ and $\frac{1}{6} \left(-9 + \sqrt{3} \sqrt{75 - 
8\mu^2} \right) $. This point always shows an acceleration and stability occurs only for $\sqrt{6} 
< \mu \leq \frac{5 \sqrt{\frac{3}{2}}}{2} $. This point is similar to $P_2$.\\

\noindent
$\mathbf{P_{11}}: (-1,0,2,0)$. $\Omega_m = 0, \Omega_r = 0, \omega_{eff} = -1 $. Eigenvalues are -5, 
-4, -4 and -1. Since $\omega_{eff} = -1$ this model behaves as a $\Lambda$CDM model. Also the eigenvectors of 
the repeated eigenvalues are independent hence this point is stable proper node. It is very interesting to 
note here that $P_{11}$ and $P_3$ are same and their stability behaviour is also same in both the cases.\\

\noindent
$\mathbf{P_{12}}: (3,0,2,0)$. $\Omega_m = -4, \Omega_r = 0, \omega_{eff} = -1 $. Eigenvalues are -4, 4, 3 and -1. 
This point is similar to $P_4$. It is shown here to note that similarly like $P_4$, $P_{12}$ is also not stable. 
So linear approximation may give correct understanding for some critical points. But as we will note later that value
of $n$ plays a crucial role.\\

\noindent
$\mathbf{P_{13}}: (4,0,2,-5)$. $\Omega_m = 0, \Omega_r = -5, \omega_{eff} = -1 $. Eigenvalues are 5, -4, 4 and 1. 
This point does have acceleration but stability does not occurs due to opposite signs of eigenvalues. This point is 
similar to $P_5$.\\

\noindent
As mentioned we will study the same critical point which were common in both the above mentioned cases. In this section, 
we will plot 3D figures to study the stability. Before that we will mention the set of autonomous differential equations
for a general case as follows:
\begin{equation}{\label{20}}
\begin{aligned}
x' &= -1 - z - 3y + x^2 - xz + w \\
y' &= \frac{xyz\mu^{2n}}{2n(2n+1)[y+(2n+1)z]} - y(2z - 4 - x) \\
z' &= - \frac{xyz\mu^{2n}}{2n(2n+1)[y+(2n+1)z]} - 2z(z-2) \\
w' &= -2zw + xw
\end{aligned}
\end{equation}
It is almost impossible to find the general eigenvalues of this system autonomous differential equations. 
Since, all the points $P_9$ to $P_{13}$ were repeated critical points, they could be considered as some of 
the many critical point of the system \eqref{20}. Then for each of the case, we do the stability analysis. 
Since here we have 2 parameters as $\mu$ and $n$ to study the stability, we first construct a 3D plot to show 
the behavior of eigenvalues for various $n$ and $\mu$. Later the region where the particular eigenvalue is negative 
could also be shown. This makes the analysis very clear.\\

\noindent
$\mathbf{P_9}:(-4,5,0,0)$. $\Omega_m = 0, \Omega_r = 0, \omega_{eff} = \frac{1}{3} $. Eigenvalues 
are -5, -4, -4 and $-\frac{2 \left(-4 n^2+\mu ^{2 n}-2 n\right)}{n (2 n+1)}$. Since $\omega_{eff} = -1$ 
this model behaves as a $\Lambda$CDM model. Also the eigenvectors of the repeated eigenvalues are independent 
hence this point is stable proper node. From the figure, we can see that this point represent stability.

\begin{center}
\includegraphics[scale=0.8]{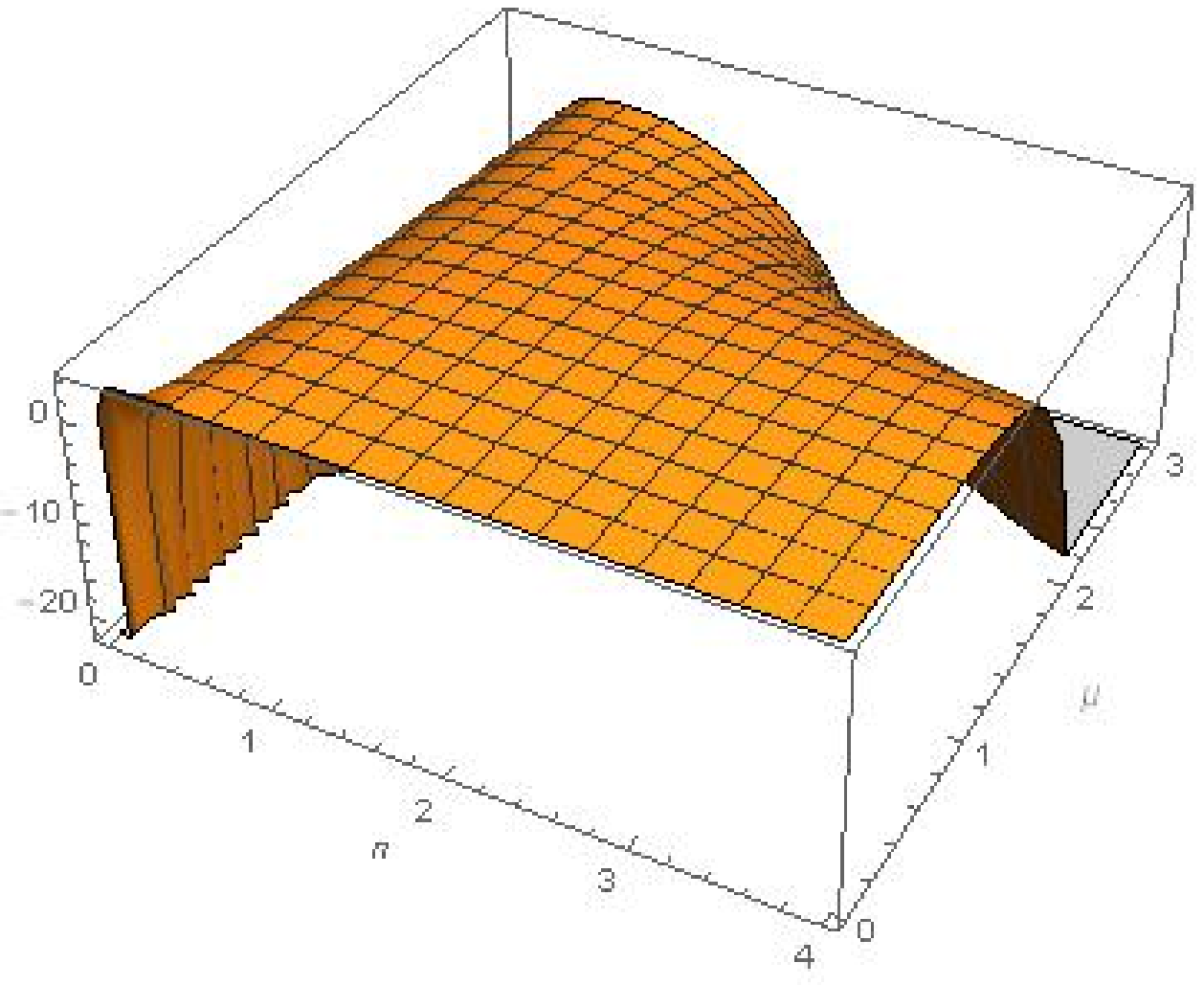}
\end{center}
We now show a 2D plot with $n$ as x-axis and $\mu$ as y-axis to show the region where the eigenvalue 
$-\frac{2 \left(-4 n^2+\mu ^{2 n}-2 n\right)}{n (2 n+1)}$ is negative. The shaded portion shows that the 
eigenvalue is negative and critical point is stable for the particular combination of $n$ and $\mu$ as soon below.

\begin{center}
\includegraphics[scale=0.8]{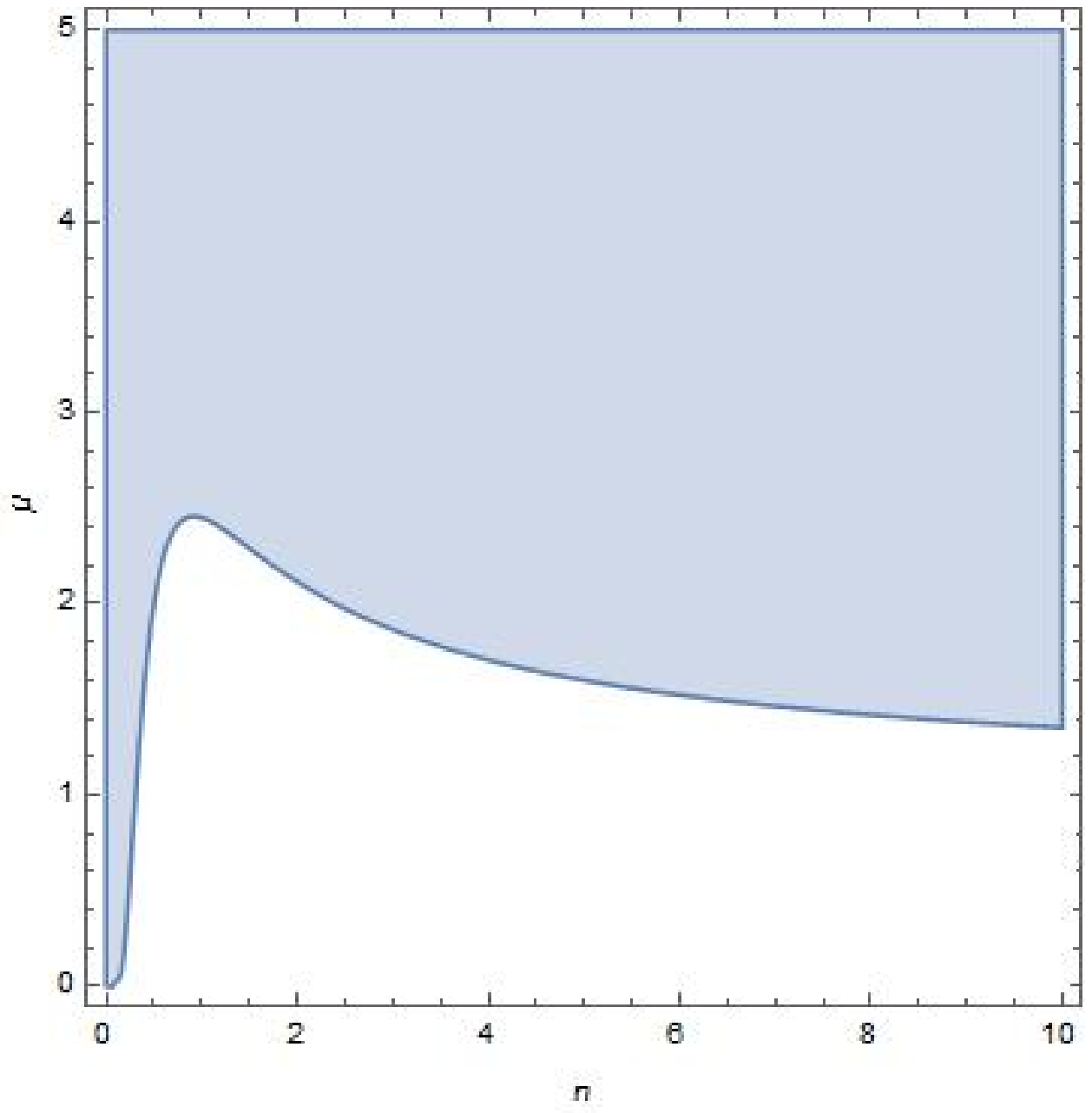}
\end{center}

\noindent
$\mathbf{P_{10}}: (0,-1,2,0)$. $\Omega_m = 0, \Omega_r = 0, \omega_{eff} = -1 $. Two of the eigenvalues 
are -3 and -4. Plot for other two is shown below which shows the stability for considered values of $\mu$ and $n$. 
In the plot yellow colour represents the real part of the eigenvalue $\frac{-3n -24 n^3-18 n^2-
\sqrt{1600 n^6+2400 n^5+1300 n^4-64 n^3 \mu ^{2 n}+300 n^3-48 n^2 \mu ^{2 n}+25 n^2-8 n \mu ^{2 n}}}{2 n (2 n+1) (4 n+1)}$ 
and blue represents the real part of eigenvalue $\frac{- 3n -24 n^3-18 n^2+\sqrt{1600 n^6+2400 n^5+1300 n^4-64 n^3 
\mu ^{2 n}+300 n^3-48 n^2 \mu ^{2 n}+25 n^2-8 n \mu ^{2 n}}}{2 n (2 n+1) (4 n+1)}$

\begin{center}
\includegraphics[scale=0.8]{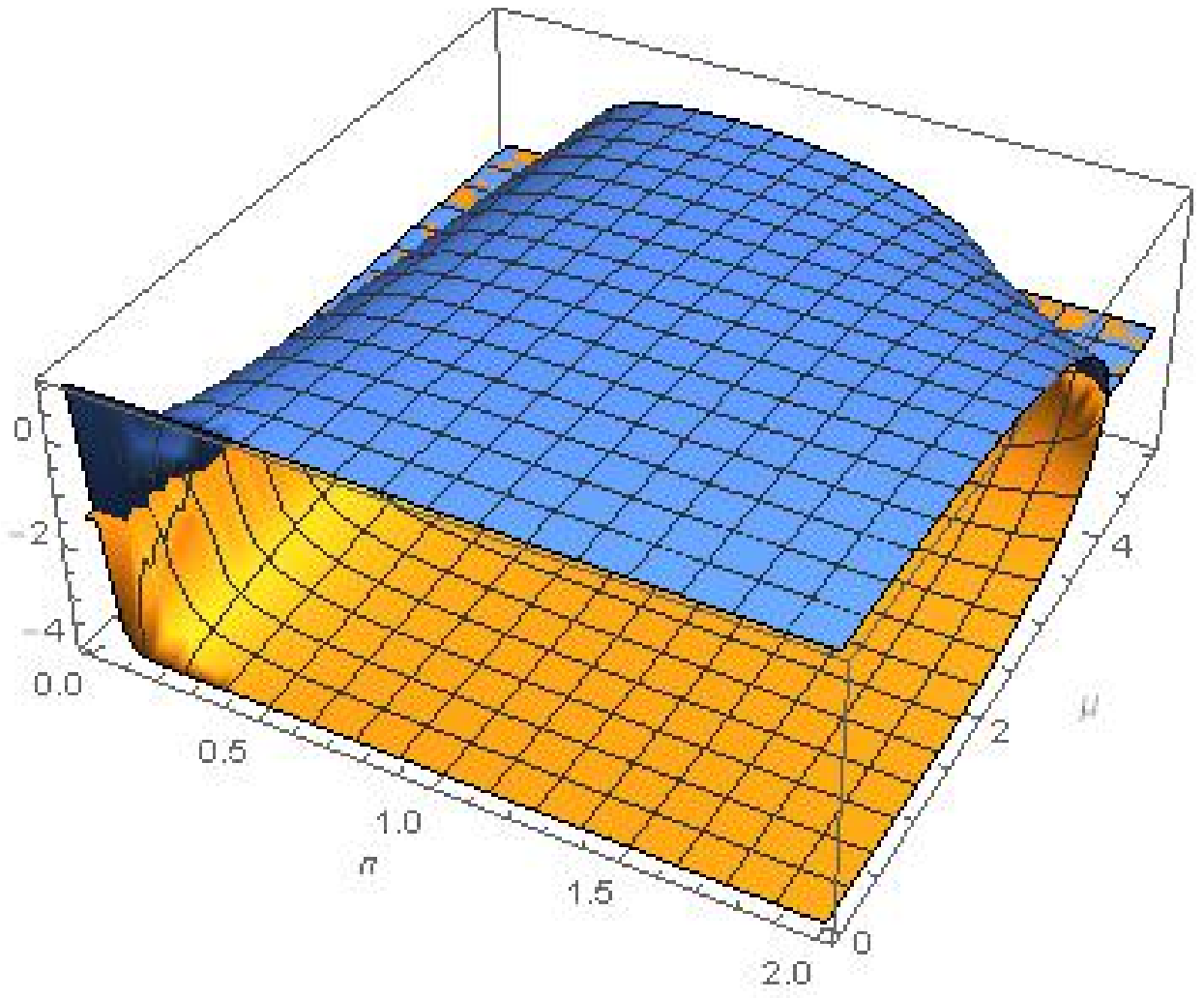}
\end{center}
We again look at the 2D plot to show the region where both the eigenvalues are negative. The shaded portion 
in the below figure represents stability.

\begin{center}
\includegraphics[scale=0.8]{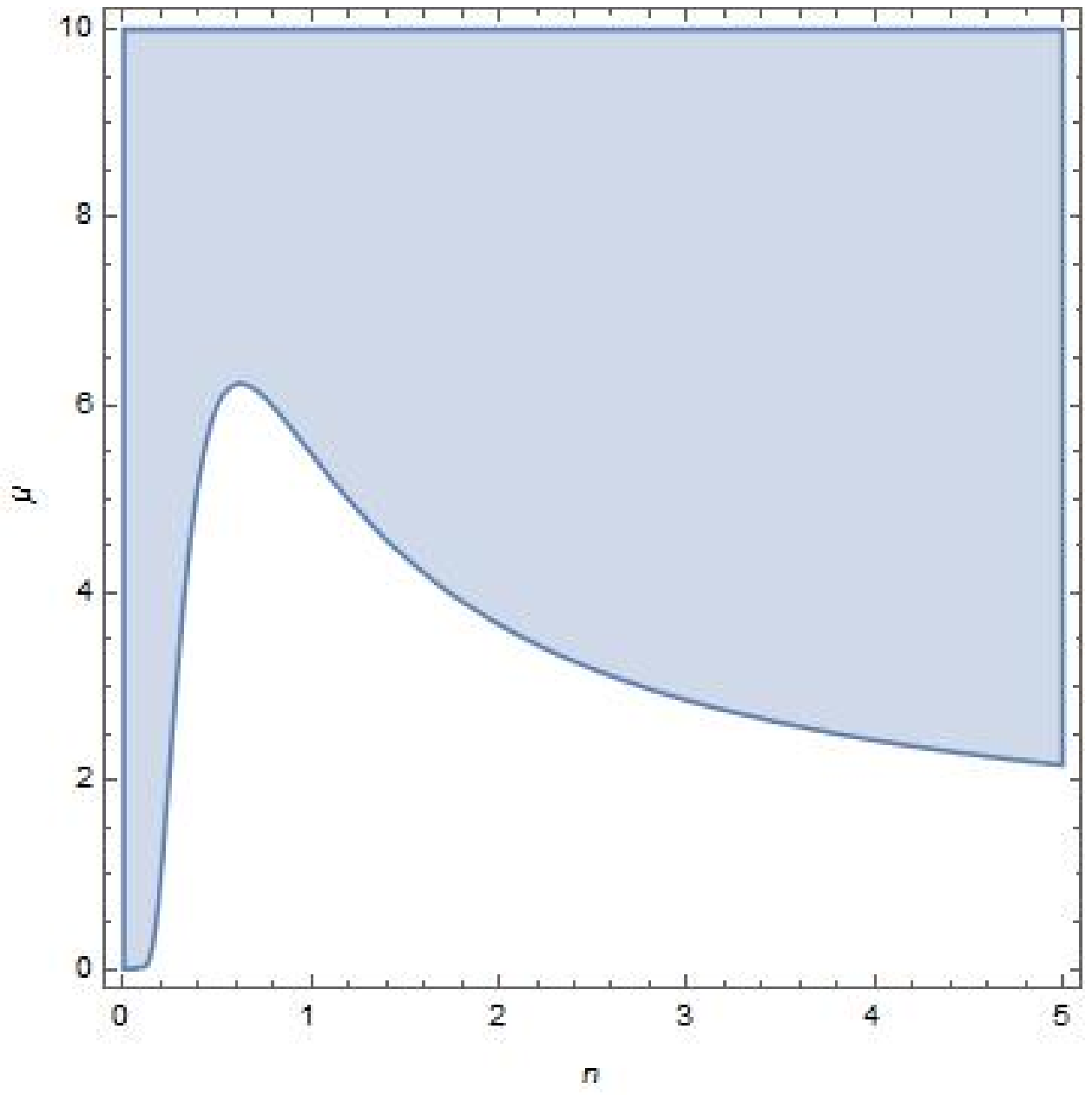}
\end{center}

\noindent
$\mathbf{P_{11}}: (-1,0,2,0)$. $\Omega_m = 0, \Omega_r = 0, \omega_{eff} = -1 $. Eigenvalues are -5, -4, -4 
and $\frac{-8 n^3-8 n^2+\mu ^{2 n}-2 n}{2 n (2 n+1)^2}$. The eigenvectors of the repeated eigenvalues are 
independent hence this point is stable proper node. We can see that the stability do occur for some combinations 
of $n$ and $\mu$. This is also evident from the 2D plot shown between $n$ and $\mu$.

\begin{center}
\includegraphics[scale=0.8]{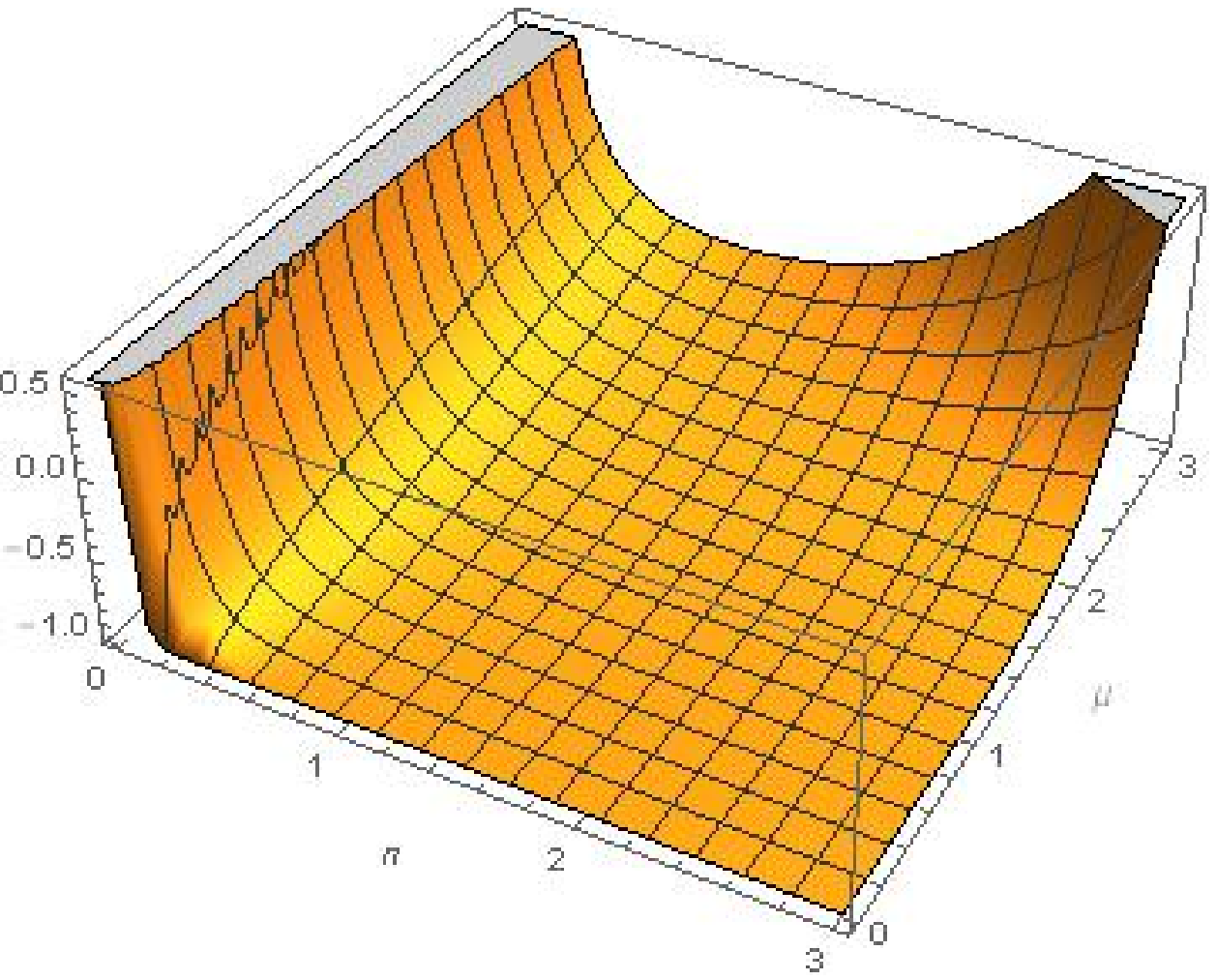}
\includegraphics[scale=0.8]{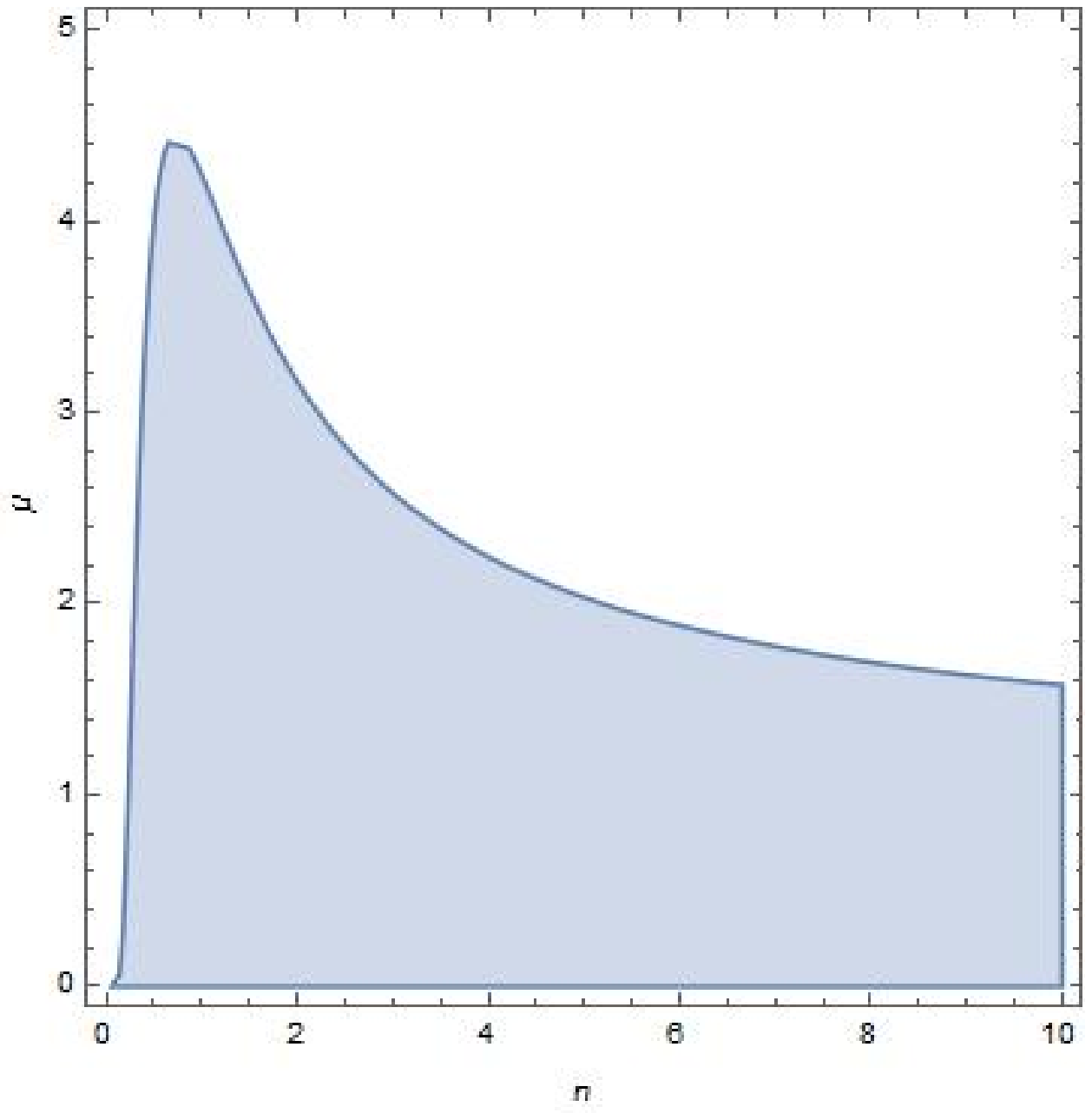}
\end{center}

\noindent
$\mathbf{P_{12}}: (3,0,2,0)$. $\Omega_m = -4, \Omega_r = 0, \omega_{eff} = -1 $. Eigenvalues are -4, 4, -1 
and $-\frac{3 \left(-8 n^3-8 n^2+\mu ^{2 n}-2 n\right)}{2 n (2 n+1)^2}$. This point is similar to $P_4$. 
It can be noted here that similarly  like $P_4$, $P_{12}$ is also not stable. Signs of eigenvalues are opposite 
hence stability could not be achieved. The plot of the eigenvalue $-\frac{3 \left(-8 n^3-8 n^2+\mu ^{2 n}-2 n\right)}
{2 n (2 n+1)^2}$ in $n - \mu$ plane is as follows: \\

\begin{center}
\includegraphics[scale=0.8]{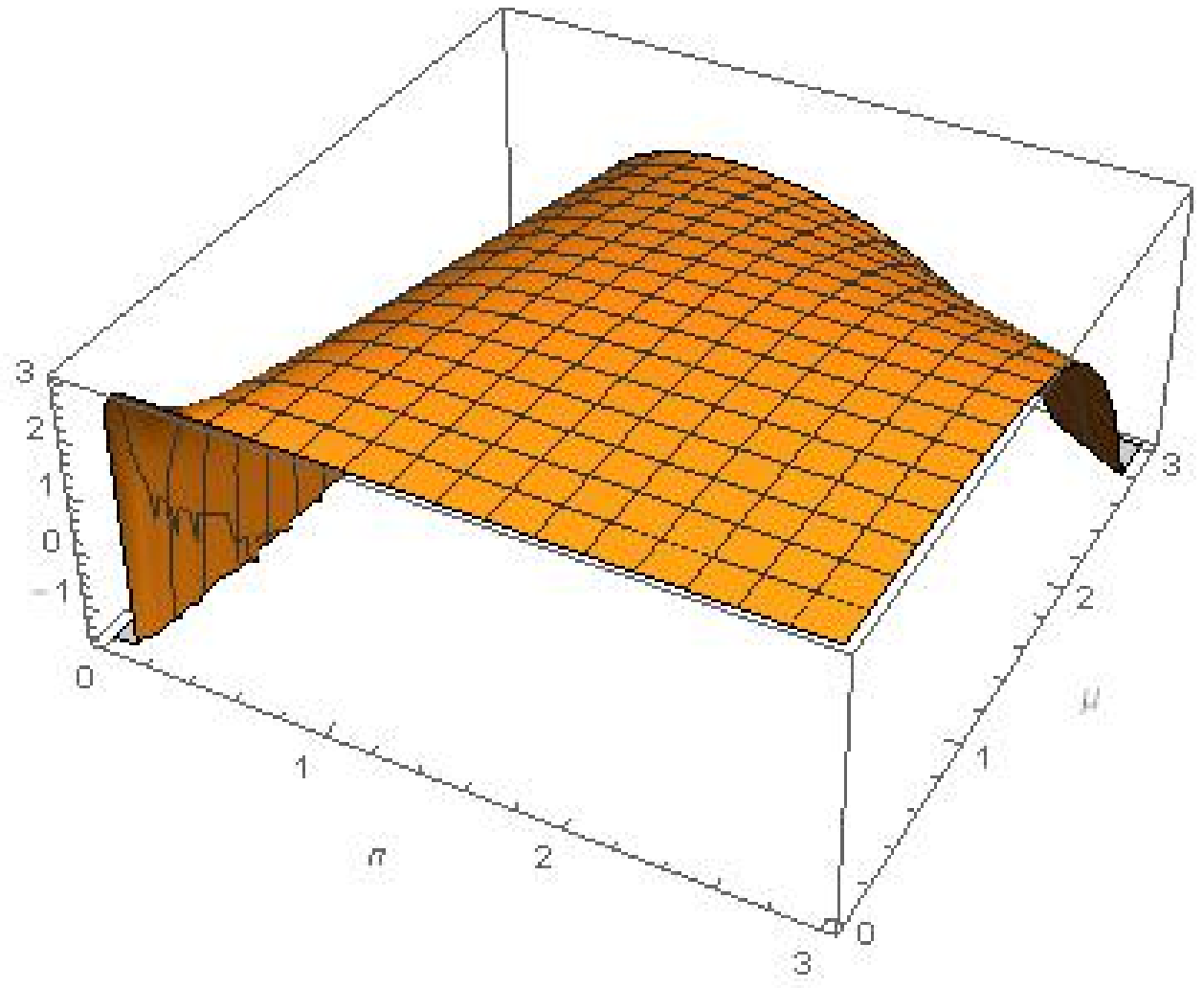}
\includegraphics[scale=0.8]{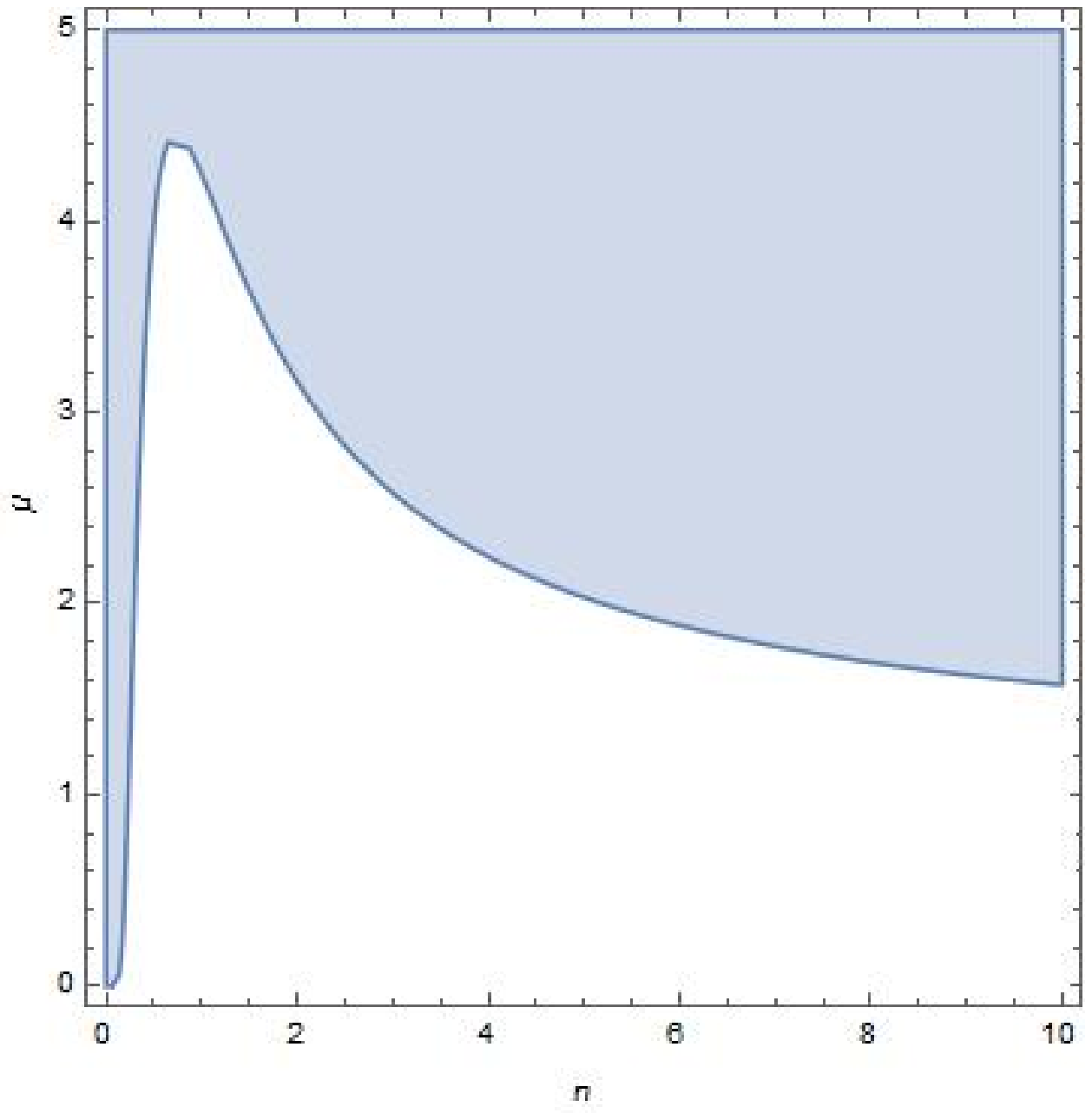}
\end{center}

\noindent
$\mathbf{P_{13}}: (4,0,2,-5)$. $\Omega_m = 0, \Omega_r = -5, \omega_{eff} = -1 $. Eigenvalues are 5, -4, 4 and 1. 
This point shows acceleration due to negative value of $\omega_{eff}$ but stability does not occurs due to opposite 
signs of eigenvalues. This point is similar to $P_5$.

\noindent
We note a peculiar thing here that whenever $x+y+z =1$, we have a stability era which are attained in $P_9, P_{10}$ 
and $P_{11}$. This indicates that the completely geometric curvature dependent universe would also bring the stability 
along with acceleration. It is seen from the observations that the dark energy is responsible for the stability of the 
universe. Here the geomtric curvature is playing the exact same role and hence giving the required stability for the 
critical points with $x+y+z = 1$. 

\section{Results and Conclusion}
In this work, the asymptotic behaviour of two different viable $f(R)$ model has been done. We first noted that
$f(R) = R - \mu R_c   \left[ 1 - (R^2/R_{c}^2)^{-n} \right]$ is an  asymptote for $f(R) = R - 
\mu R_{c} \frac{(R/R_c)^{2n}}{(R/R_c)^{2n} + 1} $ and
$f(R) = R - \mu R_{c} \left[ 1 - (1 + R^2/R_{c}^2)^{-n} \right] $ for $R \gg R_c$.
The tool of dynamical system analysis is being used for this work. It begins by introducing a set of 
dimensionless variables for the corresponding field equations. Then the system of autonomous differential equations
are formed for the $f(R)$ model under consideration. The universe is assumed to be composed only of matter and radiation
with no interaction among them. The real and compatible critical points of the system were studied.  The value 
of $\omega_{eff}$ gives the acceleration phase and signature of the eigenvalues of the Jacobian matrix of the 
corresponding critical point gives the stability analysis. This set of equations are to be solved for 4 variables. 
This work was done in two different approaches. As a part of Case-A, we fixed the value of $\mu$ and the calculations
were carried out for all the values of $n$. This resulted in formation of 8 critical point. The brief analysis of the 
result is as follows:\\
\begin{center}
\begin{tabular}{ |c|c|c| }
 \hline \textbf{Point} & \textbf{Stability} & \textbf{Acceleration}\\
 \hline
 $P_1$ & Stable for $0 < n < \frac{(\sqrt{5} - 1)}{4}$  & Never \\
 \hline
 $P_2$ &  Spiral Stable for $n \in [0.152379, 0.19908)$ & Always \\
 \hline
 $P_3$ & \shortstack{ Stable for $n > \frac{1}{24}  ( -8 + \sqrt[3]{(928 - 96 \sqrt{93}}$ \\ $+ 2 * 2^{2/3} 
 \sqrt[3]{29 + 3\sqrt{93}})$} & Always \\
 \hline
 $P_4$ & Not Stable & Always \\
 \hline
 $P_5$ & Not stable & Always \\
 \hline
 $P_6$ & Stable for few regions in $0 < n < 1.5$ & \shortstack{ $ \frac{1}{72} ( -16 + \sqrt[3]{7136 - 288 \sqrt{597}}$ \\
 + $ 2^{\frac{5}{3}}\sqrt[3]{223 + 9 \sqrt{597}} )$ \\ $< n < \frac{1}{4}(\sqrt{5} - 1)$ }\\
 \hline
 $P_7$ & Not stable & \shortstack{$\frac{1}{48} ( -12 + \sqrt[3]{3456 - 192 \sqrt{321}}$ \\ 
 $ + 4 \sqrt[3]{3 (18 + \sqrt{321}})$ \\ $< n < \frac{1}{4} (\sqrt{5} - 1)$ } \\
 \hline
 $P_8$ & \shortstack{Stable for \\ $n  > \sqrt{96 n^6 - 16 n^5 - 104 n^4 - 16 n^3 + 18 n^2 + 4 n - 1}$} & Never \\
  \hline
\end{tabular}
\end{center}

\noindent
Here, point $P_2$ is spiral stable due to the presence of some complex eigenvalues with negative real parts but
later we study that this point is indeed stable when analyzed in more general form. We now use a different approach 
as Case B, where we fix $n$ and evaluate critical points by varying $\mu$. That gives 5 real and compatible critical points.\\
\begin{center}
\begin{tabular}{ |c|c|c| }
 \hline \textbf{Point} & \textbf{Stability} & \textbf{Acceleration}\\
 \hline
 $P_9$ & Stable for $0 < \mu < \sqrt{6}$  & Never \\
 \hline
 $P_{10}$ & Stable for $\sqrt{6} < \mu \leq \frac{5 \sqrt{\frac{3}{2}}}{2} $ & Always \\
 \hline
 $P_{11}$ & Stable & Always \\
 \hline
 $P_{12}$ & Not Stable & Always \\
 \hline
 $P_{13}$ & Not stable & Always \\
 \hline
\end{tabular}
\end{center}

\noindent
The main aim of this work was to analyze the critical points of the system of \eqref{20}. It could be generalized 
that the common critical points of case A and case B are among the critical points of the system. System \eqref{20} is
a general one and does not take any particular case which makes it most suitable to study the behavior. We then looked at
the overview of these common points.

\begin{center}
\begin{tabular}{ |c|c|c| }
 \hline \textbf{Point} & \textbf{Stability} & \textbf{Acceleration}\\
 \hline
 $P_1 \sim P_9 $ & Stable for $2n(2n+1) < \mu^{2n}$  & Never \\
 \hline
 $P_2 \sim P_{10}$ & Stable for $25n(8n^2+6n+1) < 8 \mu^{2n} $ & Always \\
 \hline
 $P_3 \sim P_{11}$ & Stable for $\mu^{2n} < 2n(2n+1)^2$ & Always \\
 \hline
 $P_4 \sim P_{12}$ & Not Stable & Always \\
 \hline
 $P_5 \sim P_{13}$ & Not stable & Always \\
 \hline
\end{tabular}
\end{center}

\noindent
For these points, we showed the behavior of eigenvalue with respect to range of values of $\mu$ and $n$. Region where
the eigenvalues become negative was also shown. We noted a very peculiar property about critical points that the stability 
occurs only for the points $P_1 \sim P_9$, $P_2 \sim P_{10}$ and $P_3 \sim P_{11}$. All these points have a 
property that $x+y+z =1$. We also note a peculiar property about points $P_4 \sim P_{12}$ and $P_5 \sim P_{13}$. 
These points have $\Omega_m  =-4$ and $\Omega_r  =-5$ respectively which are highly exotic. But still $\omega_{eff}$ 
remains -1. This occurs due to the choice of the dynamical variable $x$. Also $P_4 \sim P_{12}$ and $P_5 \sim P_{13}$ 
are the only points for which the value of $x$ is positive. Here we can say that due to this positive value of $x$ 
geometric part has an anti-exotic kind of behavior and these are the only points for which stability never occurs. 
Unlike our previous work \cite{shah}, this article describes the stability analysis for a more general model. 
Moreover the stability analysis and acceleration phase were analyzed for a particular case of the model 
$f(R) = R - \mu R_{c}(R/R_{c})^{p}$ with $ 0 < p < 1, \mu, R_{c} > 0$. Whereas in this article we discuss stability 
analysis for every possible values of our parameters $\mu $ and $n$.
From all the above calculations and discussions, we could conclude that in the case of modified gravity (here $f(R)$) 
acceleration could be achieved by modifying the geometric components of the universe. Stability and acceleration phase 
are achieved which were otherwise achieved by exotic dark energy. It is obvious to note that $x+y+z =1$ means that 
geometric curvature dominated universe and we see that it is not just responsible for the acceleration of the universe 
but adding that function helps to get the stability of the universe which was otherwise achieved by dark energy. 
This work states the region where the stability of a particular point occurs. More models of such viable $f(R)$ could
be constructed to extend this work. Apart from this some different category of modification like $f(R,G)$ or Scalar 
Tensor theory could be considered and the use of dynamical system analysis could be applied further and stability 
analysis for such cases could be carried out. Scalar fields could also be considered as one of the fluid in the 
universe to study its dynamics. Scalar fields act as a dark energy component and it would be interesting to see 
the effect of both the forms of modification in the universe. Several forms of linear and non linear interactions 
in the fluids could also be studied with modification in the gravity.

\section{Acknowledgement}
Parth Shah, the author of this paper would like to acknowledge Department of Science and Technology (DST) for INSPIRE
Fellowship (Ref. No. IF160358). He also thank the Mathematics department of BITS Pilani Goa campus for providing other 
necessary research facilities. The authors are very much thankful to the anonymous reviewer for his constructive comments 
to improve the quality of work.

\end{document}